\documentclass[amsmath,amssymb,aps,prd, preprintnumbers,nofootinbib,a4paper, 11pt]{revtex4-1}
\pdfoutput=1
\usepackage{amsthm}
\usepackage{color}
\usepackage{dcolumn}
\usepackage{bm}
\usepackage{bbm}
\usepackage{pxfonts}
\usepackage{slashed} 
\usepackage{url}

\usepackage[ margin=5pt, font=normalsize,labelfont=bf,justification=raggedright]{caption}

\usepackage[bookmarks, pagebackref=false]{hyperref}
\definecolor{rossoCP3}{cmyk}{0,.88,.77,.40}
\definecolor{verdeCP3}{rgb}{0.09765625, 0.57421875, 0.1015625}
\definecolor{bluCP3}{rgb}{0, 0.23, 0.67}
 
\newcommand{\be}{\begin{eqnarray}}
\newcommand{\ee}{\end{eqnarray}}

\pdfoutput=1

 \usepackage{graphicx,hyperref}
\usepackage{float}

\newcommand{\neff}{n_{\rm eff}}
\renewcommand{\d}{{\rm d}}
\renewcommand{\bar}{\overline}

\newcommand{\R}[2]{$R_{#1#2}$}
\newcommand{\as}{$\alpha_s$}
\newcommand{\aw}{$\alpha_W$}
\newcommand{\nlojet}{\textsc{NLOJet++}}
\newcommand{\beq}{\begin{eqnarray}}
\newcommand{\eeq}{\end{eqnarray}}
\newcommand{\pTonetwo}{\left<p_{T1,2}\right>}

\begin{document}
\title{\Large  \color{rossoCP3} ~~  Challenging Asymptotic Freedom\footnote{Invited talk at the CERN workshop on: "High precision $\alpha_s$ measurements: from LHC to FCC-ee", October 12th -13th, 2015,  Geneva, Switzerland.  }  }
 \author{Francesco  Sannino}
\email{sannino@cp3-origins.net} 
\affiliation{{\color{rossoCP3} CP$^{3}$-Origins} \& Danish Institute for Advanced Study {\color{rossoCP3} DIAS}, University of Southern Denmark, Campusvej 55, DK-5230 Odense M, Denmark}

\begin{abstract}
Several extensions of the standard model feature new colored states that besides modifying the running of the QCD coupling could even lead to the loss of asymptotic freedom. Such a loss would potentially diminish the Wilsonian fundamental value of the theory.  However, the recent discovery of complete asymptotically safe  vector-like theories \cite{Litim:2014uca}, i.e. featuring an interacting UV fixed point in all couplings, elevates these theories to a fundamental status and opens the door to alternative UV completions of (parts of) the standard model. If, for example, QCD rather than being asymptotically free  becomes asymptotically safe there would be consequences on the early time evolution of the Universe (the QCD plasma would not be free).  It is therefore important to test, both directly and indirectly, the strong coupling running at the highest possible energies. I will review here the attempts made in  \cite{Becciolini:2014lya} to use pure QCD observables at the Large Hadron Collider (LHC) to place bounds on new colored states. Such bounds do not depend on the detailed properties of the new hypothetical states  but on their effective number  and mass.  We will see that these direct  constraints cannot exclude a potentially safe, rather than free, QCD asymptotic nature. A  safe QCD scenario would imply that  quarks and gluons are only approximately free at some intermediate energies, otherwise they are always in chains.  
\\[.1cm]
{\footnotesize  \it Preprint: CP$^3$-Origins-2015-049 DNRF90, DIAS-2015-49.}
 \end{abstract}

\maketitle

\thispagestyle{empty}
 
\section{The need to test QCD at higher energies}
  The standard model (SM) of particle interactions is an extremely successful theory at and below the Fermi scale.  This scale is   identified with the spontaneous breaking of the electroweak symmetry.  
 
The  mathematical structure of the (SM) contains a gauge sector associated to local invariance of the semi-simple group $SU(3)\times SU(2)\times U(1)$.  As soon as an elementary scalar sector is introduced new accidental interactions emerge. These are the Yukawa interactions responsible for the flavour research program, and the Higgs scalar self interactions\footnote{The Higgs self-coupling is known when assuming the minimal SM Higgs realisation but it has not yet been directly measured.}. Accidental interactions are not associated to a gauge principle and their number and type is limited by global symmetries and the request of renormalisability. In four dimensions accidental symmetries are associated, at the tree level, to either relevant operators -- from the infrared physics point of view -- such as the Higgs mass term, or to marginal operators such as Yukawa interactions.  Gauge sectors, on the other hand, lead only to marginal operators. These theories are known as Gauge-Yukawa theories and, especially after the discovery of the Higgs, it has become imperative to acquire a deeper understanding of their dynamics. 

One can further classify Gauge-Yukawa theories according to whether they admit  UV complete (in all the couplings) fixed points. The presence of such a fixed point guarantees the fundamentality of the theory since, setting aside gravity, it means that the theory is valid at arbitrary short distances.

\subsection*{Is QCD asymptotic free above the Fermi scale?} 
If the UV fixed point occurs for vanishing values of the couplings the interactions become asymptotically free in the UV  \cite{Gross:1973id,Politzer:1973fx}. The fixed point is approached logarithmically and therefore, at short distances, perturbation theory can be used. Asymptotic freedom is an UV phenomenon that still allows for several possibilities in the IR, depending on the specific underlying theory \cite{Sannino:2009za}. At low energies, for example, another interacting fixed point can occur. In this case the theory displays both long and short distance conformality. However the theory is interacting at short distances and the IR spectrum of the theory is continuous \cite{Georgi:2007ek}. Another possibility that can occurr in the IR is that a dynamical mass is generated leading to either confinement or chiral symmetry breaking, or both.

QCD does not possess an interacting IR fixed point because it generates dynamically a mass scale that can be, for example, read off any non-Goldstone hadronic state such as the nucleon or the vector meson $\rho$. Because, however, we have measured the strong coupling only up to subTeV energies (see the {\it World summary of 
$\alpha_s$ (2015)} contribution by S. Bethke) one cannot yet experimentally infer that QCD is asymptotically free.  

In fact it is intriguing to explore both theoretical and experimental extensions of the SM in which  QCD looses asymptotic freedom at short distances. For example, asymptotic freedom can be lost by considering additional vector-like colored matter at or above the Fermi scale.

\subsection*{Alternative  safe QCD scenario}

 On the other hand, loosing asymptotic freedom would, at one loop level, unavoidably lead to the emergence of a Landau pole. A result that potentially diminishes the Wilsonian fundamental value of the theory. 
  
 There is, however, another largely unexplored possibility. Namely that an UV interacting fixed point is re-instated either perturbatively or non-perturbatively. This equally interesting and safe UV completion of strong interactions would have far-reaching consequences when searching for UV complete extensions of the SM as well as cosmology.  If experimentally true, in fact, it would radically change our view of fundamental interactions and profoundly affects our understanding of the early cosmological evolution of the universe (this is so since the primordial plasma would not be free at high temperatures). 

That such an interacting UV fixed point can exists for nonsupersymmetric vector-like theory it has been recently established in \cite{Litim:2014uca,Litim:2015iea}. Furthermore, no additional symmetry principles such as space-time supersymmetry \cite{Bagger:1990qh} are required to ensure well-defined and predictive theories in the UV  \cite{Litim:2011cp}. Instead, the fixed point arises dynamically through  renormalisable interactions between non-Abelian gauge fields, fermions, and scalars, and in a regime where asymptotic freedom is absent.
The potentially dangerous growth of the gauge coupling towards the UV is countered by Yukawa interactions, while the Yukawa and scalar couplings are tamed by the fluctuations of  gauge and fermion fields. This has led to  theories with ``complete asymptotic safety'', meaning interacting UV  fixed points in all couplings \cite{Litim:2014uca}.
This is quite distinct  from the conventional setup of
``complete asymptotic freedom'' \cite{Gross:1973ju,Cheng:1973nv,Callaway:1988ya}, where the UV dynamics of Yukawa and scalar interactions is  brought under control by asymptotically free gauge fields; see \cite{Holdom:2014hla,Giudice:2014tma} for recent studies.

It is also straightforward to engineer QCD-like IR behaviour in the theory investigated in \cite{Litim:2014uca}, including confinement and chiral symmetry breaking. In practice one  decouples, at sufficiently high energies, the unwanted fermions by adding mass terms or via spontaneous symmetry breaking in such a way that at lower energies the running of the gauge coupling mimics QCD. The use in \cite{Litim:2014uca} of the Veneziano limit of large number of colors and flavors was instrumental to prove the existence of the UV fixed point in all couplings of the theory within perturbation theory.  Tantalising indications that such a fixed point exists nonperturbatively, and without the need of elementary scalars, appeared in \cite{Pica:2010xq}, and they were further explored in  \cite{Shrock:2013cca,Litim:2014uca}. Nonperturbative techniques are needed to establish the existence of such a fixed point when the number of colors and flavours is taken to be three and the number of UV light flavours is large but finite. 

Interestingly the supersymmetric cousins of the theory investigated in  \cite{Litim:2014uca} (technically super QCD with(out) a meson-like chiral superfield), once asymptotic freedom is lost,  cannot be asymptotically safe \cite{Martin:2000cr,Intriligator:2015xxa}. The results were further generalised  and tested versus a-maximisation \cite{Intriligator:2015xxa}. 

\begin{figure}
	\centering
	\includegraphics[width=0.95\linewidth]{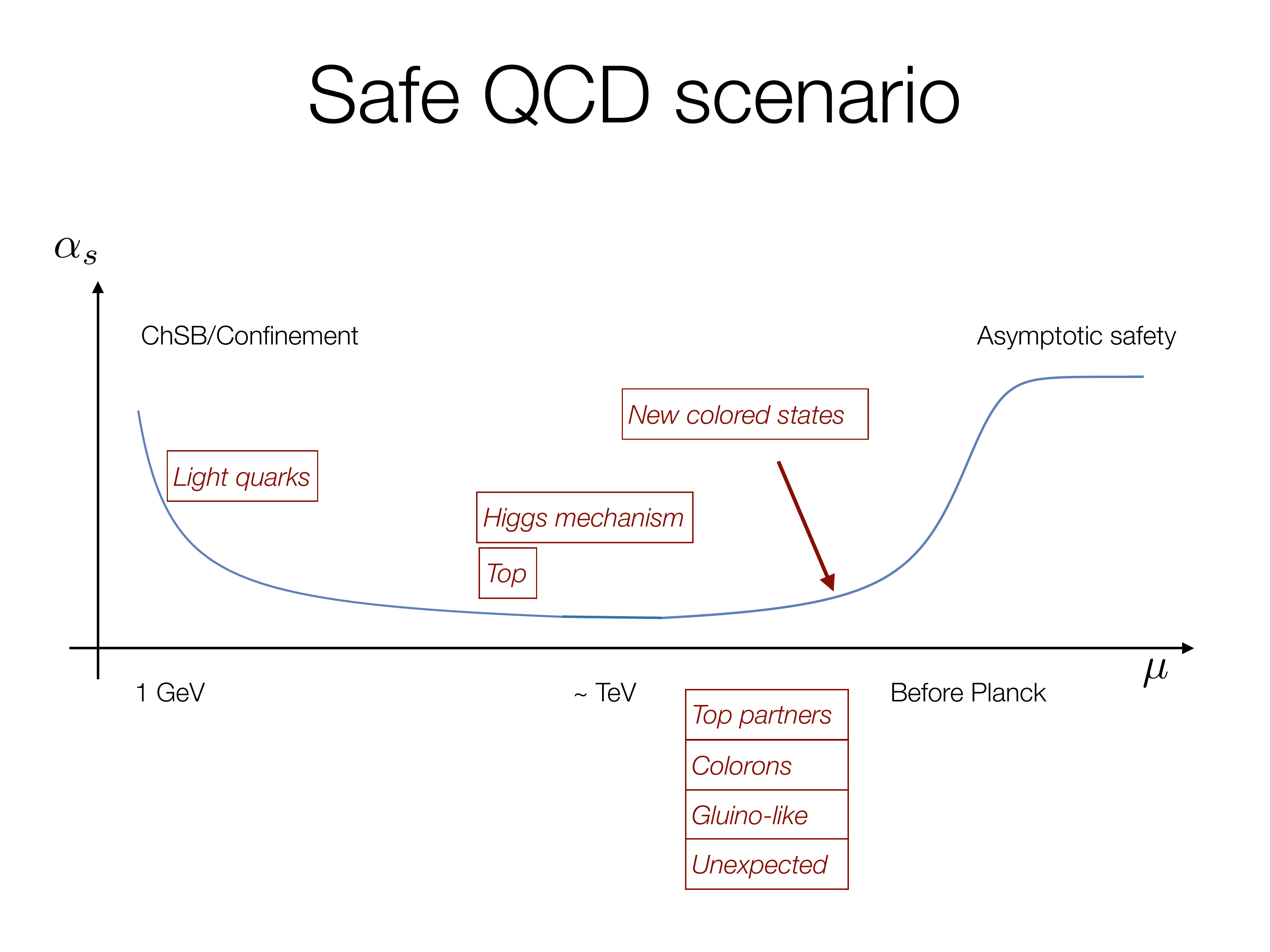}
	\caption{Asymptotically safe scenario of QCD expressed in terms of the running of \as.}
	\label{safeQCD}
\end{figure}

One can envision several extensions of the SM that can lead to rapid changes in the running of the QCD coupling at and above the Fermi scale. Since I have shown that  asymptotic freedom can be traded for asymptotic safety while leaving the fundamental properties of QCD untouched, it becomes crucial to test the high energy behaviour of strong interactions. In Figure~\ref{safeQCD} I present a cartoon version of how the QCD running coupling could look when going from the IR to the UV. Of course the final asymptotically safe value, to be reached below the Planck scale, does not have to be large. 

Although it would be extremely interesting to consider indirect constraints coming from cosmological and/or high energy astrophysical observations ranging from cosmic rays to compact objects  one can investigate direct constraints coming from future (and current) LHC experiments. These, as we shall see, can help setting model-independent bounds on the effective number of new colored states  around the Fermi scale. 

  \section{Constraining new colored matter at the LHC}
  
The LHC experiments are, in fact, already probing the evolution of the strong coupling \as\ up to the TeV scale. It was showed in \cite{Becciolini:2014lya} how the ratio of 3- to 2-jets cross sections is affected by the presence of new physics and argued that it can be used to place model-independent bounds on new particles carrying QCD color charge. It was also argued that such states need to be heavier than a few hundred GeVs. Stronger constraints can be derived model by model but, in this case, the results are not general.

 In~\cite{Chatrchyan:2013txa} the first determination of the strong coupling $\alpha_s (M_Z)$ from measurements of momentum scales beyond 0.6 TeV was presented.
This determination was performed studying the behaviour of the ratio \R32 of the inclusive 3-jet cross section 
to the 2-jet cross section, defined in greater detail below. The result is in agreement with the world average value of $\alpha_s (M_Z)$.\footnote{Even more recent measurements of \as\ at high energy scales have appeared after completion of this project, see~\cite{Khachatryan:2014waa} or other contributions to this report.}

In \cite{Becciolini:2014lya} it has been argued that it is possible to constrain the presence of new colored states using such a measurement that probes quantum chromodynamics (QCD) at harder scales.

It was also pointed out that such constraints should be taken with the grain of salt because of concerns regarding the validity of the interpretation given in the experimental analyses which warrants further studies.

Rather than duelling on the validity of the experimental analysis  the focus in \cite{Becciolini:2014lya}  has been on the large potential value of such an observable for placing bounds on new physics beyond the SM.  In the process one gains insight related to the presence of new colored particles. For example, it was shown that their effect on the parton distribution functions is negligible. This is true, at least, when constructing ratios of cross-sections. Therefore in the absence of clear final states observables it was shown that the presence of new colored particles appears directly in the running of \as.
  
This approach provides complementary information with respect to typical direct limits, where several assumptions are made to specify production and decay of a  given new colored particle.
For instance if the new particles have the required quantum numbers, searches for di-jet resonances are particularly constraining~\cite{Han:2010rf}, while there are models evading these bounds for which the results presented here may be relevant~\cite{Kubo:2014ova}.
Furthermore the impact of \as running only depends on the mass of the new states and on their color representation (and number). It is in this sense that an exclusion bound from such a measurement is, to a good approximation, model independent.

Efforts to constrain light colored states in the same spirit  appeared, for example, in~\cite{Berger:2004mj,Berger:2010rj}. Here the effects of a gluino-like state on the global analysis of scattering hadron data were considered, while in~\cite{Kaplan:2008pt}  model-independent bounds on new colored particles were derived using event shape data from the LEP experiments. 

Finally, this type of approach generalises to other sectors of the SM, and the electroweak sector could for instance be constrained from  measurements of Drell-Yan processes at higher energies ~\cite{Alves:2014cda}.
 
 Observables involving a low inclusive number of hard jets constitute ideal candidates to test  QCD at the highest possible energy scales and therefore we focus on the ratio of 3- to 2-jets (differential) cross sections, \R32 \cite{Arnison:1985zm, Appel:1985iv, Abe:1995rw, Abbott:2000ua, ATLAS:2013lla, Chatrchyan:2013txa}.

Following CMS ~\cite{Chatrchyan:2013txa}:
\beq
	R_{32}\left( \pTonetwo \right) \equiv \frac{\d\sigma^{n_j\geq3} / \d\pTonetwo}{\d\sigma^{n_j\geq2} / \d\pTonetwo},
\eeq
where $\pTonetwo$ is the average transverse momentum of the two leading jets in the event,
\beq
	\pTonetwo \equiv \frac{p_{T1} + p_{T2}}{2}.
\eeq
Other choices are possible regarding the kinematic variable \cite{ATLAS:2013lla}.

In the original work \cite{Becciolini:2014lya}  we considered the CMS analysis based on 5~fb$^{-1}$ of data collected at 7~TeV centre-of-mass energy~\cite{Chatrchyan:2013txa}.
Jets were defined requiring transverse momenta of at least 150~GeV and rapidities less than 2.5, using the anti-kT algorithm~\cite{Cacciari:2008gp} with size parameter R = 0.7 and E-recombination scheme.

The computations for inclusive multijet cross sections include the next-to-leading order corrections in \as~and \aw~\cite{Ellis:1992en, Nagy:2001fj, Moretti:2006ea, Dittmaier:2012kx}.%
\footnote{Recent progress making use of new unitarity-based techniques will allow for complete NNLO results in a near future~\cite{Ridder:2013mf, Currie:2013dwa}.}
NLO QCD corrections are implemented in \nlojet~\cite{Nagy:2003tz}, that allows to evaluate the 3- and 2-jets cross sections at the parton-level within the Standard Model.

The problem lies in the definition of the factorisation and the renormalisation scales identified with $\pTonetwo$ in the theoretical calculations presented by CMS. Since 3-jet events
involve multiple scales, this simplified assignment may not represent the dynamics in play appropriately enough to allow a straightforward interpretation of the experimental data as a measurement of \as\ at $\pTonetwo$;
the observable may be mainly sensitive to the value of the strong coupling at some fixed lower scale.
Although the ideas presented here hinge on a resolution of this issue, finding the proper redefinition or reinterpretation of \R32 goes beyond the original scope of \cite{Becciolini:2014lya}.
 
Hypothetical new colored particles can contribute to \R32 through a modification of the running of \as\ and of the PDFs, and as additional contributions to the partonic cross section at leading or next-to-leading order. It was argued in \cite{Becciolini:2014lya}  that the most important of these effects is the change in \as\ and that the correspondence between \R32 and  the strong coupling constant is reliable, even in the presence of new physics. The reader will find in \cite{Becciolini:2014lya} an in depth description of the validity and caveats of the approach. 

In general, new states may contribute at tree-level to the jet cross sections if their quantum numbers allow it.
Although this would lead to important modifications in \R32 these  contributions may partially cancel, the same way NLO corrections do as shown in the \cite{Becciolini:2014lya}.
 Notice that the colored states are stable, a new, heavy fermion in the final state could in principle be misidentified as a jet, since it would hadronize and end somewhere in the detector. There are however stringent constraints on the existence of such bound states~\cite{Aad:2013gva}.
 
 Let's now consider the running of \as when new colored fermions appear at high energies  and therefore write the associated $\beta$ function \cite{Becciolini:2014lya}
 \beq
	\beta(\alpha_s) \equiv \mu \frac{\partial \alpha_s}{\partial \mu} = - \frac{\alpha_s^2}{2 \pi} \left( b_0 + \frac{\alpha_s}{4 \pi} \, b_1 + \ldots \right),
\label{eq:beta}
\eeq
then the coefficients $b_0$ and $b_1$ in any mass-independent renormalisation scheme read
\beq
	b_0 & = & 11 - \frac{2}{3} n_f - \frac{4}{3} n_X T_X, \\
	b_1 & = & 102 - \frac{38}{3} n_f - 20 \, n_X T_X \left( 1 + \frac{C_X}{5} \right),
\eeq
where $n_f$ is the number of quark flavours (i.e. $n_f = 6$ at scales $Q > m_t$), $n_X$ the number of new (Dirac) fermions, and $T_X$ and $C_X$ group theoretical factors depending in which representation of the color group the new fermions transform. 
 
 Since the adjoint representation is real --- like the gluino in the MSSM --- a Majorana mass term can be written for a single Weyl fermion, and $n_X$ can take half-integer values. At leading order, the modification in the running of \as\ only depends on a single parameter $\neff \equiv 2 n_X T_X$, counting the effective number of new fermions
\beq
	\neff = n_{\mathbf{3}\oplus \bar{\mathbf{3}}} + 3 \, n_{\mathbf{8}}
		+ 5 \, n_{\mathbf{6}\oplus \bar{\mathbf{6}}}
		+ 15 \, n_{\mathbf{10}\oplus \bar{\mathbf{10}}},
	\label{eq:neff}
\eeq
where $n_{\mathbf{3}\oplus \bar{\mathbf{3}}}$, $n_{\mathbf{6}\oplus \bar{\mathbf{6}}}$ and $n_{\mathbf{10}\oplus \bar{\mathbf{10}}}$ are the number of new Dirac fermions in the triplet, sextet and decuplet representations respectively, and $n_{\mathbf{8}}$ the number of Weyl fermions in the adjoint representation. Asymptotic freedom is lost for $\neff > 10.5$. In view of our initial motivation we do not restrict ourselves to asymptotically free theories.

Furthermore, one Dirac fermion corresponds to four complex scalar degrees of freedom;
scalar particles in the spectrum thus contribute to $\neff$ four times less than corresponding Dirac fermions.
For instance, the full content of the Minimal Supersymmetric Standard Model (1 adjoint Weyl fermion and 12 fundamental complex scalars) counts as $\neff = 6$.

\begin{figure}
	\centering
	\includegraphics[width=0.7\linewidth]{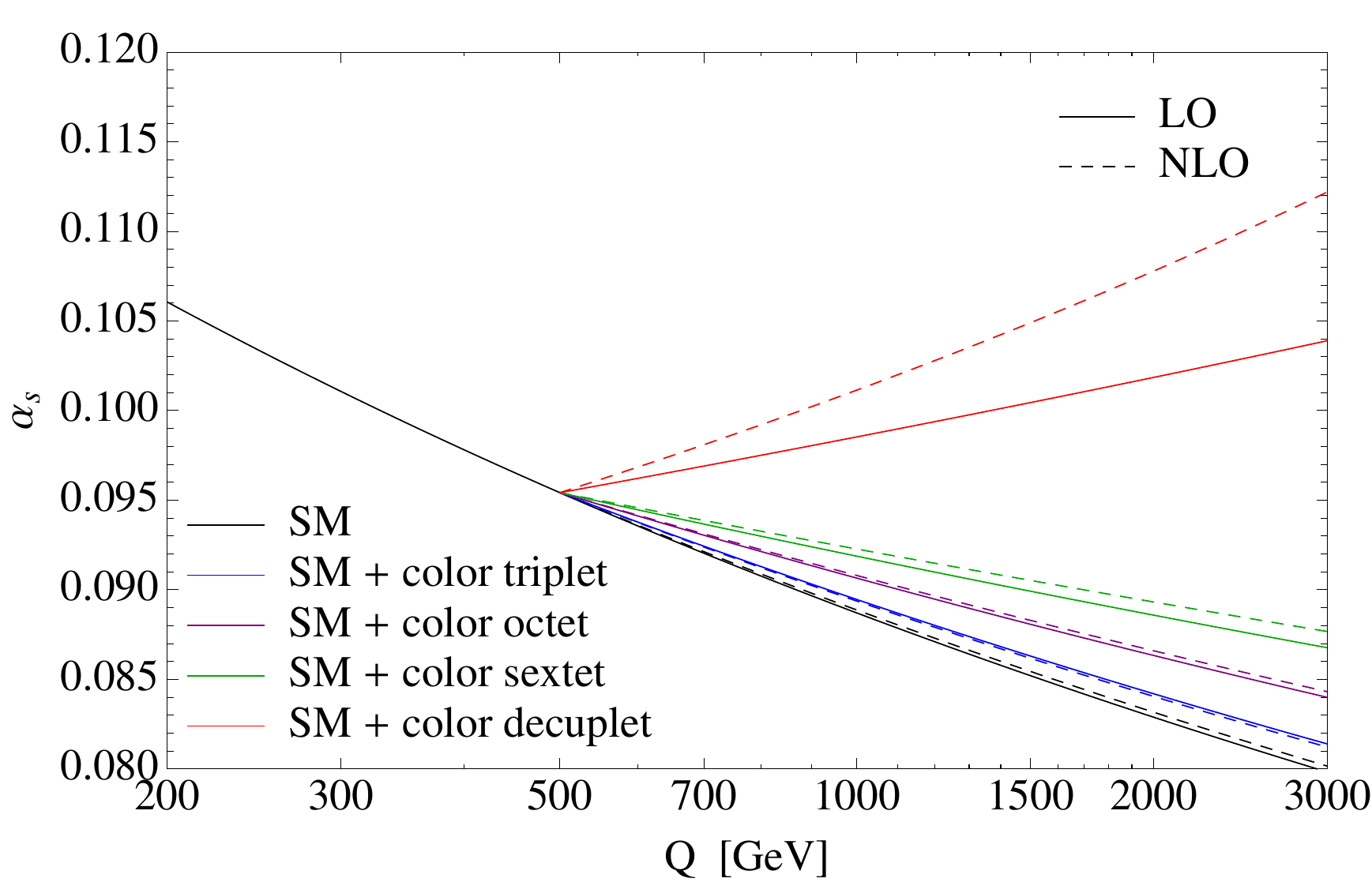}
	\caption{Example of the change in \as\ induced by a new fermion of mass 500~GeV in various representations of the color gauge group. The running of \as\ is performed at NLO, showing for comparison the running at LO from the mass of the new fermion.}
	\label{fig:alpha}
\end{figure}
The running of \as\ as given by the $\beta$ function above is only valid at energies larger than the mass of the new colored fermions --- for simplicity, we assumed that they all have the same mass $m_X$ and that they are heavier than the top quark.
One can now match \as\ between the high-energy regime and the effective description without the new fermions  at  $m_X$.

The relative change in \as\ induced by fermions in various representation can be assessed from Fig.~\ref{fig:alpha}.  It is also useful to introduce the approximate expression \cite{Becciolini:2014lya}:
\beq
	\frac{\alpha_s(Q)}{\alpha_s^{SM}(Q)} \approx  1 + \frac{\neff}{3\pi} \alpha_s(m_X)
				\log\left( \frac{Q}{m_X} \right),
	\nonumber\\ \textrm{~for~} Q \geq m_X,
	\label{eq:alphaapprox}
\eeq
where $\alpha_s^{SM}(Q)$ is the Standard Model value of the running coupling.

The reader will find the detailed investigation of the impact of the new colored fermions on the PDFs  in \cite{Becciolini:2014lya}.  It is sufficient here to say that, at the precision level of the analysis, the modification of PDFs can thus be safely neglected for \R32.

To illustrate the exclusion potential of high-scale measurements of \as\ the bounds were presented for $\neff$ depending on the scale of new physics $m_X$, and using CMS~\cite{Chatrchyan:2013txa}.

\begin{table}
\centering
\begin{tabular}{c|c}
$Q$ [GeV] & $\alpha_s^{exp}(Q) \pm \sigma(Q)$ \\
\hline
474 & $0.0936 \pm 0.0041$ \\
664 & $0.0894 \pm 0.0031$ \\
896 & $0.0889 \pm 0.0034$ \\
\end{tabular}
\caption{High-scale determinations of \as\ from measurements of \R32 by CMS~\cite{Chatrchyan:2013txa}.}
\label{tab:alpha}
\end{table}
We assume the CMS estimates of \as\ reproduced in Tab.~\ref{tab:alpha}) and further add to the analysis the world average measurement of the strong coupling $\alpha_s(M_Z) = 0.1185 \pm 0.0006$~\cite{Beringer:1900zz}; since its uncertainty is much smaller than the ones of the other data points,
we take as fixed input $\alpha_s(M_Z) = 0.1185$.

The induced probability measure over the parameter-space to constrain is proportional to  \cite{Becciolini:2014lya}
\beq
\exp\left[{-\frac{1}{2} \sum_Q \left(\frac{\alpha_s^{exp}(Q) - \alpha_s^{th}(Q ; \neff, m_X)}{\sigma(Q)}\right)^2}\right] 
\times \text{priors}, \quad
\eeq
where $\alpha_s^{th}(Q ; \neff, m_X)$ is the theoretical prediction for the value of the strong coupling at the scale $Q$, which is a function of $\neff$ and $m_X$.

The theoretical predictions for \as\ are obtained by running up to $Q$ from the $Z$-mass at two-loop order, as described in eq.~\eqref{eq:beta}, which is sufficient for our purpose.
Beyond leading-order, $\neff$ is not enough to parametrise the importance of new physics effects:
the quadratic Casimir $C_X$ needs to be specified.
In \cite{Becciolini:2014lya} its valued was varied between $4/3$ and $6$ --- the values corresponding to fermions in the fundamental or decuplet representations, respectively --- to demonstrate that it has little relevance.

\begin{figure}
	\centering
		\includegraphics[width=0.7\linewidth]{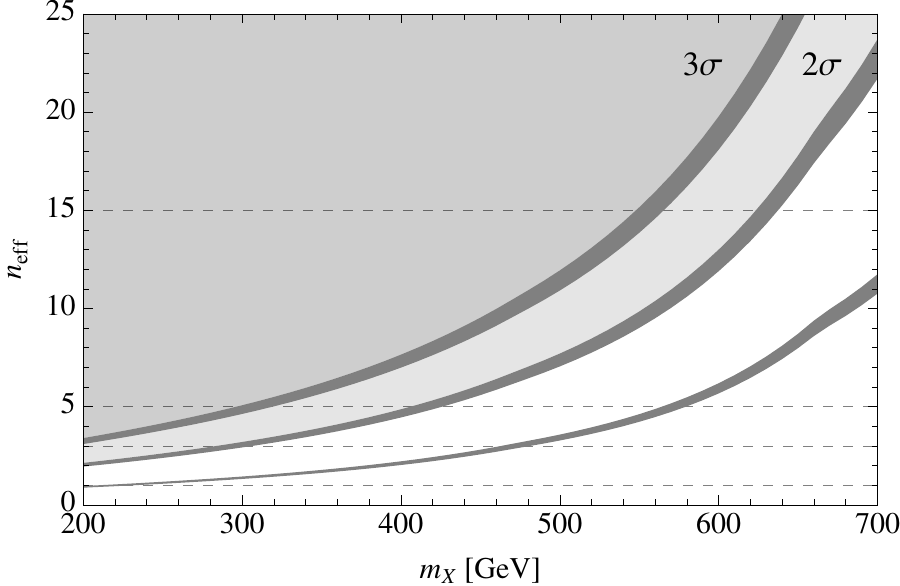}
	\caption{The shaded regions indicate the upper bounds on $\neff$ at $2\sigma$ and $3\sigma$ confidence levels, assuming the scale of new physics $m_X$ is known.
	They are delimited by grey bands whose width show the effect of varying the Casimir $C_X$.
	 As further indication, the third band shows a $1\sigma$ limit.
	To guide the eye, the dashed horizontal lines indicate values of $\neff$ corresponding to one fundamental, one adjoint, one two-index symmetric and one three-index symmetric fermion (see eq.~\eqref{eq:neff}).}
\label{fig:exclusion}
\end{figure}

Assuming that the mass of the new states is known  the upper bound  on $\neff$  is shown in Fig.~\ref{fig:exclusion}.
It is useful, following \cite{Becciolini:2014lya}, to represent the exclusion potential of currently available experimental data as shown in Tab.~\ref{tab:limits}.

\begin{table}
\centering
\begin{tabular}{c|c|c}
 color content & $\neff$ & $m_X$ in GeV \\
\hline
Gluino & 3 & 280 \\
Dirac sextet & 5 & 410 \\
MSSM & 6  & 450 \\
Dirac decuplet & 15 & 620 \\
\end{tabular}
\caption{
$95\%$ CL mass exclusion bounds for various values of $\neff$ according to a toy-analysis of the latest CMS measurement of \R32~\cite{Chatrchyan:2013txa}.
}
\label{tab:limits}
\end{table}

\section{Conclusions}
I first motivated the need to test QCD beyond the asymptotically free paradigm and then argued, based on the findings in  \cite{Becciolini:2014lya},  that pure QCD observables can help placing interesting bounds on new physics beyond the Standard Model. Such bounds are insensitive to the detailed properties of the new hypothetical states and mainly depend on their effective number $\neff$ and mass. Although these limits on colored particles might not be the most stringent for any specific model, are nevertheless unavoidable because of  their model-independent nature. As an explicit example the ratio of 3- to 2-jets inclusive differential cross sections \R32 has been discussed along with the associated caveats \cite{Becciolini:2014lya}. 

With LHC running at almost double the centre-of-mass energy one expects that higher mass exclusion bounds will be available. The relative simplicity of the analysis suggested in  \cite{Becciolini:2014lya} allows to swiftly extract limits on new physics  as new data becomes available. 

It is clear from the above that direct  constraints cannot exclude a potentially safe, rather than free, QCD asymptotic behavior. If the alternative safe QCD scenario were true quarks and gluons are never entirely free.

\acknowledgments
It is a pleasure for me to thank Celine Boehm, Richard Brower, Lance Dixon, Ken Intriligator, Klaus Rabbertz and Natascia Vignaroli for interesting discussions. I am indebted to Diego Becciolini, Marc Gillioz, Ken Intriligator, Daniel Litim, Matin Mojaza, Marco Nardecchia, Claudio Pica and Michael Spannowsky for collaborating  on some of the results and ideas partially summarised here.  I learnt a great deal from collaborating with them. I would like to note that the idea of a potential UV completion of QCD that is asymptotically safe rather than free is, to the best of my knowledge, original and it is not contained in references \cite{Litim:2014uca,Becciolini:2014lya}. Last but not the least I thank David d'Enterria and Peter Z. Skands for having organised this topical and relevant workshop.


\begin{thebibliography}{99}



\bibitem{Litim:2014uca} 
  D.~F.~Litim and F.~Sannino,
  JHEP {\bf 1412}, 178 (2014)
  doi:10.1007/JHEP12(2014)178
  [arXiv:1406.2337 [hep-th]].


\bibitem{Becciolini:2014lya} 
  D.~Becciolini, M.~Gillioz, M.~Nardecchia, F.~Sannino and M.~Spannowsky,
  Phys.\ Rev.\ D {\bf 91}, no. 1, 015010 (2015)
  [Phys.\ Rev.\ D {\bf 92}, no. 7, 079905 (2015)]
  doi:10.1103/PhysRevD.91.015010, 10.1103/PhysRevD.92.079905
  [arXiv:1403.7411 [hep-ph]].




\bibitem{Gross:1973id} 
  D.~J.~Gross and F.~Wilczek,
  Phys.\ Rev.\ Lett.\  {\bf 30}, 1343 (1973).
  doi:10.1103/PhysRevLett.30.1343


\bibitem{Politzer:1973fx} 
  H.~D.~Politzer,
  Phys.\ Rev.\ Lett.\  {\bf 30}, 1346 (1973).
  doi:10.1103/PhysRevLett.30.1346

\bibitem{Sannino:2009za} 
  F.~Sannino,
  Acta Phys.\ Polon.\ B {\bf 40}, 3533 (2009)
  [arXiv:0911.0931 [hep-ph]].

\bibitem{Georgi:2007ek} 
  H.~Georgi,
  Phys.\ Rev.\ Lett.\  {\bf 98}, 221601 (2007)
  doi:10.1103/PhysRevLett.98.221601
  [hep-ph/0703260].

\bibitem{Litim:2015iea} 
  D.~F.~Litim, M.~Mojaza and F.~Sannino,
  arXiv:1501.03061 [hep-th].

\bibitem{Bagger:1990qh} 
  J.~Bagger and J.~Wess,
  JHU-TIPAC-9009.


\bibitem{Litim:2011cp} 
  D.~F.~Litim,
  Phil.\ Trans.\ Roy.\ Soc.\ Lond.\ A {\bf 369}, 2759 (2011)
  [arXiv:1102.4624 [hep-th]].


\bibitem{Gross:1973ju} 
  D.~J.~Gross and F.~Wilczek,
  Phys.\ Rev.\ D {\bf 8}, 3633 (1973).
  doi:10.1103/PhysRevD.8.3633


\bibitem{Cheng:1973nv} 
  T.~P.~Cheng, E.~Eichten and L.~F.~Li,
  Phys.\ Rev.\ D {\bf 9}, 2259 (1974).
  doi:10.1103/PhysRevD.9.2259


\bibitem{Callaway:1988ya} 
  D.~J.~E.~Callaway,
  Phys.\ Rept.\  {\bf 167}, 241 (1988).
  doi:10.1016/0370-1573(88)90008-7


\bibitem{Holdom:2014hla} 
  B.~Holdom, J.~Ren and C.~Zhang,
  JHEP {\bf 1503}, 028 (2015)
  doi:10.1007/JHEP03(2015)028
  [arXiv:1412.5540 [hep-ph]].


\bibitem{Giudice:2014tma} 
  G.~F.~Giudice, G.~Isidori, A.~Salvio and A.~Strumia,
  JHEP {\bf 1502}, 137 (2015)
  doi:10.1007/JHEP02(2015)137
  [arXiv:1412.2769 [hep-ph]].


\bibitem{Pica:2010xq} 
  C.~Pica and F.~Sannino,
  Phys.\ Rev.\ D {\bf 83}, 035013 (2011)
  doi:10.1103/PhysRevD.83.035013
  [arXiv:1011.5917 [hep-ph]].

\bibitem{Shrock:2013cca} 
  R.~Shrock,
  Phys.\ Rev.\ D {\bf 89}, no. 4, 045019 (2014)
  doi:10.1103/PhysRevD.89.045019
  [arXiv:1311.5268 [hep-th]].

\bibitem{Martin:2000cr} 
  S.~P.~Martin and J.~D.~Wells,
  Phys.\ Rev.\ D {\bf 64}, 036010 (2001)
  doi:10.1103/PhysRevD.64.036010
  [hep-ph/0011382].


\bibitem{Intriligator:2015xxa} 
  K.~Intriligator and F.~Sannino,
  JHEP {\bf 1511}, 023 (2015)
  doi:10.1007/JHEP11(2015)023
  [arXiv:1508.07411 [hep-th]].


\bibitem{Chatrchyan:2013txa} 
  S.~Chatrchyan {\it et al.} [CMS Collaboration],
  Eur.\ Phys.\ J.\ C {\bf 73}, no. 10, 2604 (2013)
  doi:10.1140/epjc/s10052-013-2604-6
  [arXiv:1304.7498 [hep-ex]].


\bibitem{Khachatryan:2014waa} 
  V.~Khachatryan {\it et al.} [CMS Collaboration],
  Eur.\ Phys.\ J.\ C {\bf 75}, no. 6, 288 (2015)
  doi:10.1140/epjc/s10052-015-3499-1
  [arXiv:1410.6765 [hep-ex]].


\bibitem{Han:2010rf} 
  T.~Han, I.~Lewis and Z.~Liu,
  JHEP {\bf 1012}, 085 (2010)
  doi:10.1007/JHEP12(2010)085
  [arXiv:1010.4309 [hep-ph]].


\bibitem{Kubo:2014ova} 
  J.~Kubo, K.~S.~Lim and M.~Lindner,
  Phys.\ Rev.\ Lett.\  {\bf 113}, 091604 (2014)
  doi:10.1103/PhysRevLett.113.091604
  [arXiv:1403.4262 [hep-ph]].


\bibitem{Berger:2004mj} 
  E.~L.~Berger, P.~M.~Nadolsky, F.~I.~Olness and J.~Pumplin,
  Phys.\ Rev.\ D {\bf 71}, 014007 (2005)
  doi:10.1103/PhysRevD.71.014007
  [hep-ph/0406143].


\bibitem{Berger:2010rj} 
  E.~L.~Berger, M.~Guzzi, H.~L.~Lai, P.~M.~Nadolsky and F.~I.~Olness,
  Phys.\ Rev.\ D {\bf 82}, 114023 (2010)
  doi:10.1103/PhysRevD.82.114023
  [arXiv:1010.4315 [hep-ph]].


\bibitem{Kaplan:2008pt} 
  D.~E.~Kaplan and M.~D.~Schwartz,
  Phys.\ Rev.\ Lett.\  {\bf 101}, 022002 (2008)
  doi:10.1103/PhysRevLett.101.022002
  [arXiv:0804.2477 [hep-ph]].

 

\bibitem{Alves:2014cda} 
  D.~S.~M.~Alves, J.~Galloway, J.~T.~Ruderman and J.~R.~Walsh,
  JHEP {\bf 1502}, 007 (2015)
  doi:10.1007/JHEP02(2015)007
  [arXiv:1410.6810 [hep-ph]].


\bibitem{Arnison:1985zm} 
  G.~Arnison {\it et al.} [UA1 Collaboration],
  Phys.\ Lett.\ B {\bf 158}, 494 (1985).
  doi:10.1016/0370-2693(85)90801-9


\bibitem{Appel:1985iv} 
  J.~A.~Appel {\it et al.} [UA2 Collaboration],
  Z.\ Phys.\ C {\bf 30}, 341 (1986).
  doi:10.1007/BF01557598


\bibitem{Abe:1995rw} 
  F.~Abe {\it et al.} [CDF Collaboration],
  Phys.\ Rev.\ Lett.\  {\bf 75}, 608 (1995).
  doi:10.1103/PhysRevLett.75.608


\bibitem{Abbott:2000ua} 
  B.~Abbott {\it et al.} [D0 Collaboration],
  Phys.\ Rev.\ Lett.\  {\bf 86}, 1955 (2001)
  doi:10.1103/PhysRevLett.86.1955
  [hep-ex/0009012].


\bibitem{ATLAS:2013lla} 
  [ATLAS Collaboration],
  ATLAS-CONF-2013-041.


\bibitem{Cacciari:2008gp} 
  M.~Cacciari, G.~P.~Salam and G.~Soyez,
  JHEP {\bf 0804}, 063 (2008)
  doi:10.1088/1126-6708/2008/04/063
  [arXiv:0802.1189 [hep-ph]].


\bibitem{Ellis:1992en} 
  S.~D.~Ellis, Z.~Kunszt and D.~E.~Soper,
  Phys.\ Rev.\ Lett.\  {\bf 69}, 1496 (1992).
  doi:10.1103/PhysRevLett.69.1496


\bibitem{Nagy:2001fj} 
  Z.~Nagy,
  Phys.\ Rev.\ Lett.\  {\bf 88}, 122003 (2002)
  doi:10.1103/PhysRevLett.88.122003
  [hep-ph/0110315].


\bibitem{Moretti:2006ea} 
  S.~Moretti, M.~R.~Nolten and D.~A.~Ross,
  Nucl.\ Phys.\ B {\bf 759}, 50 (2006)
  doi:10.1016/j.nuclphysb.2006.09.028
  [hep-ph/0606201].


\bibitem{Dittmaier:2012kx} 
  S.~Dittmaier, A.~Huss and C.~Speckner,
  JHEP {\bf 1211}, 095 (2012)
  doi:10.1007/JHEP11(2012)095
  [arXiv:1210.0438 [hep-ph]].


\bibitem{Ridder:2013mf} 
  A.~Gehrmann-De Ridder, T.~Gehrmann, E.~W.~N.~Glover and J.~Pires,
  Phys.\ Rev.\ Lett.\  {\bf 110}, no. 16, 162003 (2013)
  doi:10.1103/PhysRevLett.110.162003
  [arXiv:1301.7310 [hep-ph]].


\bibitem{Currie:2013dwa} 
  J.~Currie, A.~Gehrmann-De Ridder, E.~W.~N.~Glover and J.~Pires,
  JHEP {\bf 1401}, 110 (2014)
  doi:10.1007/JHEP01(2014)110
  [arXiv:1310.3993 [hep-ph]].


\bibitem{Nagy:2003tz} 
  Z.~Nagy,
  Phys.\ Rev.\ D {\bf 68}, 094002 (2003)
  doi:10.1103/PhysRevD.68.094002
  [hep-ph/0307268].


\bibitem{Aad:2013gva} 
  G.~Aad {\it et al.} [ATLAS Collaboration],
  Phys.\ Rev.\ D {\bf 88}, no. 11, 112003 (2013)
  doi:10.1103/PhysRevD.88.112003
  [arXiv:1310.6584 [hep-ex]].


\bibitem{Beringer:1900zz} 
  J.~Beringer {\it et al.} [Particle Data Group Collaboration],
  Phys.\ Rev.\ D {\bf 86}, 010001 (2012).
  doi:10.1103/PhysRevD.86.010001
\end{thebibliography}
\end{document}